\documentclass[conference]{IEEEtran}
\IEEEoverridecommandlockouts

\usepackage{cite}
\usepackage{amsmath,amssymb,amsfonts}
\usepackage{algorithm,algorithmic}
\usepackage{graphicx,flushend,multirow,booktabs,subfigure}
\usepackage{textcomp}
\def\BibTeX{{\rm B\kern-.05em{\sc i\kern-.025em b}\kern-.08em
    T\kern-.1667em\lower.7ex\hbox{E}\kern-.125emX}}
\begin{document}

\title{Deep unrolled primal dual network for TOF-PET list-mode image reconstruction}

\author{Rui Hu, Chenxu Li, Kun Tian, Jianan Cui, Yunmei Chen and Huafeng Liu
\thanks{This work is supported in part by the Talent Program of Zhejiang Province (No: 2021R51004), by the National Natural Science Foundation of China (No: U1809204), by the Key Research and Development Program of Zhejiang Province (No: 2021C03029), and by NSF grants: DMS2152961.}
\thanks{Corresponding author: liuhf@zju.edu.cn (Huafeng Liu).}
\thanks{Rui Hu, Chenxu Li Kun Tian and Huafeng liu are with State Key Laboratory of Modern Optical Instrumentation, 
College of Optical Science and Engineering, Zhejiang University, Hangzhou, 
China.(e-mail: rickhu@zju.edu.cn; 22060538@zju.edu.cn; 22230106@zju.edu.cn; jianancui@zjut.edu.cn; yun@math.ufl.edu; liuhf@zju.edu.cn}
\thanks{Jianan Cui is with Institute of Information Processing and Automation, College of Information Engineering, Zhejiang University of Technology, Hangzhou 310001, China}
\thanks{Yunmei Chen is with Department of Mathematics, University of Florida, Gainesville, FL 32611 USA}
\thanks{Huafeng Liu is also with Jiaxing Key Laboratory of Photonic Sensing \& Intelligent Imaging, Jiaxing 314000, China and with Intelligent Optics \& Photonics Research Center, Jiaxing Research Institute, Zhejiang University, Jiaxing 314000, China}}

\maketitle

\begin{abstract}
 Time-of-flight (TOF) information provides more accurate location data for annihilation photons, thereby enhancing the quality of PET reconstruction images and reducing noise. List-mode reconstruction has a significant advantage in handling TOF information. However, current advanced TOF PET list-mode reconstruction algorithms still require improvements when dealing with low-count data. Deep learning algorithms have shown promising results in PET image reconstruction. Nevertheless, the incorporation of TOF information poses significant challenges related to the storage space required by deep learning methods, particularly for the advanced deep unrolled methods. In this study, we propose a deep unrolled primal dual network for TOF-PET list-mode reconstruction. The network is unrolled into multiple phases, with each phase comprising a dual network for list-mode domain updates and a primal network for image domain updates. We utilize CUDA for parallel acceleration and computation of the system matrix for TOF list-mode data, and we adopt a dynamic access strategy to mitigate memory consumption. Reconstructed images of different TOF resolutions and different count levels show that the proposed method outperforms the list-mode ordered subset expectation maximization (LM-OSEM), total-variation regularized list-mode expectation maximization (LM-EMTV), list-mode stochastic primal dual hybrid gradient (LM-SPDHG), total variation regularized stochastic primal dual hybrid gradient (LM-SPDHG-TV) and FastPET method in both visually and quantitative analysis. These results demonstrate the potential application of deep unrolled methods for TOF-PET list-mode data and show better performance than current mainstream TOF-PET list-mode reconstruction algorithms, providing new insights for the application of deep learning methods in TOF list-mode data. The codes for this work are available at https://github.com/RickHH/LMPDnet

\end{abstract}


\section{Introduction}
\label{sec:introduction}
\IEEEPARstart{T}{ime}-of-flight (TOF) positron emission tomography (PET) is an emerging imaging technique that has garnered significant attention in recent years~\cite{karp2008benefit,surti2020update,qi2022positronium}. By measuring the time difference between the emission of two annihilation photons, TOF PET provides additional information about the location of the annihilation event, leading to improved spatial resolution and image quality compared to conventional PET. This increased sensitivity is particularly beneficial in cancer imaging, where detecting small lesions and early-stage disease is critical for effective treatment~\cite{jakoby2011physical,mehranian2022deep,sanaat2022deep}. 

However, incorporating TOF information poses significant computational challenges for image reconstruction~\cite{kadrmas2009impact}, especially for sinogram data. List-mode data acquisition and reconstruction methods have emerged as a promising alternative for TOF PET image reconstruction, offering various advantages over sinogram-based methods~\cite{hu2006dynamic,wang2006systematic,zhou2014efficient,cui2011fully,zhang2018optimization}. List-mode reconstruction focuses on processing only detected events, enhancing computational efficiency. Additionally, list-mode reconstruction is well-suited for dynamic gated imaging\cite{shi2019direct} or short frame imaging \cite{spangler2021ultra} and demonstrates superior adaptability in managing irregularities in scanner geometry and patient motion\cite{jin2013list}.

The prevailing list-mode reconstruction algorithms for TOF-PET employ iterative methods such as Maximum Likelihood Expectation Maximization (MLEM)~\cite{huesman2000list,reader2002accelerated}, Ordered Subset Expectation Maximization (OSEM)~\cite{pratx2006fully}, and the more recent Primal-Dual Hybrid Gradient (PDHG)~\cite{schramm2022fast}. Despite their advantages, these approaches are hindered by high noise levels in low count data, significant computational load, and low reconstruction speeds. Regularization methods, such as Total Variation (TV)~\cite{raczynski20203d,zhang2016preliminary} and Non Local Mean (NLM)~\cite{wang2015anatomical}, have been employed to improve image quality of these iterative methods. However, determining the appropriate regularization parameters remains a challenge, often requiring extensive experiments and expertise\cite{zhang2017regularization}. 

\begin{figure*}   
\centering
\includegraphics[width=\linewidth]{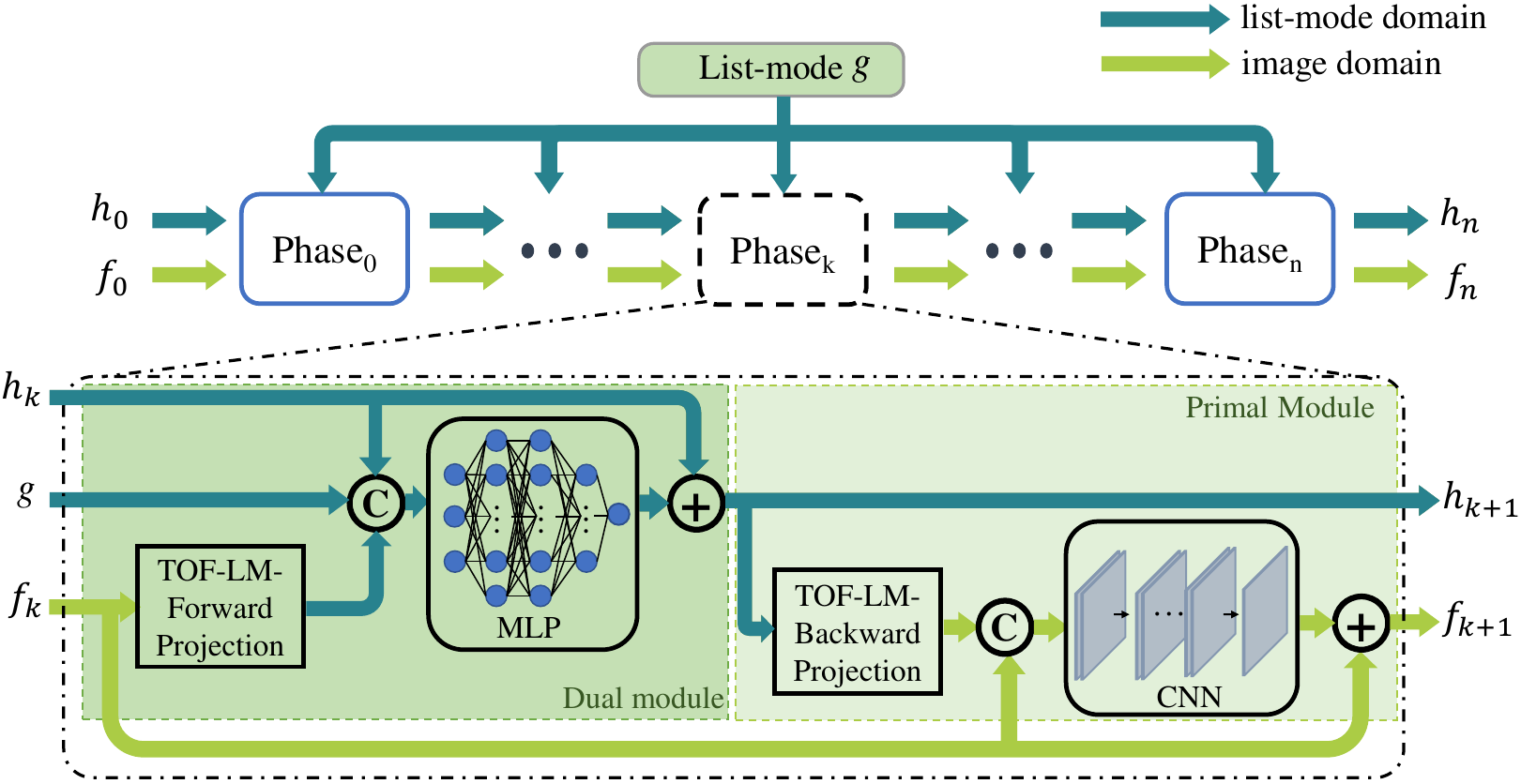}  
\caption{The reconstruction scheme of the proposed LMPDnet. The reconstruction process is unrolled with $n$ phases, each phase contains one dual module for dual variable of list-mode domain updating and one primal module for primal variable of image domain updating. The teal arrows indicate the path of the list-mode data, and the lime green arrows indicate the path of the image domain data.}
\label{Fig1}
\end{figure*}

Recently, deep learning has emerged as a popular approach for PET image reconstruction~\cite{reader2020deep,zaharchuk2019next,gong2019machine}, owing to its high computational efficiency and data-driven characteristics. Several studies have demonstrated the effectiveness of deep learning techniques in improving PET image reconstruction accuracy and reducing noise, artifacts, and overall image degradation. Among these techniques, the model-based deep learning (MoDL)~\cite{aggarwal2018modl,mehranian2020model,lim2020improved,gong2019mapem,hu2022transem} has shown promising results with good interpretability, making it a promising direction for PET image reconstruction. However, these methods face limitations in handling TOF information effectively. This is partly due to the substantial storage burden imposed by TOF information on sinogram data, which hinders the training of MoDL methods. Additionally, the current mainstream neural networks used in MoDL methods, such as convolutional neural networks (CNN), struggle to process binary list-mode data. 

Whiteley et al. introduced FastPET~\cite{whiteley2020fastpet}, an innovative approach leveraging deep learning for TOF-PET list-mode reconstruction. The method involved transforming the initial list-mode data into a histogram before inputting it into the neural network for learning, allowing the network to facilitate the conversion between the histogram and image domains. Kibo and Fumio further enhanced FastPET by incorporating additional directional information to improve image quality~\cite{ote2022deep}. However, these techniques require converting the original list-mode data to a histogram in the image domain, rather than directly utilizing list-mode data with TOF information. Li et al. presented BP-Net~\cite{lv2022back}, a novel approach for list-mode reconstruction in TOF-PET that employed back projection of list-mode data to facilitate the conversion to the image domain. However, this conversion might result in an unwanted reduction in both spatial and temporal sampling. LM-DIPRecon~\cite{ote2023list} integrated the list-mode dynamic row action maximum likelihood algorithm (LM-DRAMA)~\cite{tanaka2003subset} with deep image prior, achieving direct reconstruction from list-mode data with enhanced image quality. Notably, this method did not incorporate TOF information.

In this paper, we propose a novel deep unrolled model-based learning method for TOF-PET list-mode reconstruction. The forward and backward projection of list-mode data are embedded in the network. The entire reconstruction process is unrolled into several phases, with each phase including a dual module for list-mode domain updates and a primal module for image domain updates. The rest of this paper is organized as follows. Section \textrm{II} introduces the network structure of proposed method and details of the TOF-PET list-mode reconstruction process. Section \textrm{III} describes the simulations, dataset, and the experimental results, followed by further discussion of the results in Section \textrm{IV}. Section \textrm{V} provides the final conclusions.

\section{Methods}
\subsection{TOF-PET list-mode reconstruction problem}
In TOF-PET list-mode reconstruction, the measured list-mode data is represented as a sequence of TOF bin indices that identify the coincidence event in a specific line of response (LOR)~\cite{ote2023list}:
\begin{equation}
g = \{i(t)|t = 1,2,...,N\}
\end{equation}
where $i(t)$ is an index of TOF bin measuring the $t$-th event, $N$ is the total number of the events. 

% where $x \in {\mathbb{R}^{M}}$ denotes tracer activity distribution image, $M$ is the number of pixels in the image. $P \in {\mathbb{R}^{N \times M}}$ denotes the list-mode TOF forward projection model, which also called the system response matrix including the attenuation factor, normalization and TOF kernel modeling. Vector $s \in {\mathbb{R}^{N}}$ is the contribution of the random and scattered coincidences.

Maximum the penalized Poisson log-likelihood function is a common way solving the PET list-mode image reconstruction problem:
\begin{equation}
\mathop {\arg \max }\limits_{x \geqslant 0} L(g|x) + \beta R(x)
\end{equation}
where the $R(\cdot)$ denotes the regularization term, $\beta$ is the regularization parameter and $L(g|x)$ denotes the list-mode log-likelihood function which is given by:
\begin{equation}
\begin{split}
L(g \mid x)  =\sum_{t=1}^N \log \sum_{j=1}^M a_{i(t) j} x_j-\sum_{i=1}^I \sum_{j=1}^M a_{i j} x_j \\
 =\sum_{t=1}^N \log (A x)_{i(t)}-\sum_{i=1}^I(A x)_i
\end{split}
\end{equation}
where $A \in {\mathbb{R}^{N \times M}}$ denotes the list-mode TOF forward projection model, which also called the system response matrix including the attenuation factor, normalization and TOF kernel modeling. The $a_{ij}$ represents the contribution of voxel $j$ to the TOF bin $i$. $I$ is the total number of TOF bins. $M$ is the number of pixels.

% 介绍一下Learned Primal Dual如何解决上述优化问题
% 应用在list-mode数据上的困难：1. event数据难以用卷积神经网络进行处理 2. list-mode投影问题：不能实时计算、网络反向传播困难 3.
\subsection{Learned Primal Dual}  
Previously, the learned primal dual (LPD) method~\cite{adler2018learned} used convolutional neural networks to replace the proximal operators in the primal dual hybrid gradient (PDHG) algorithm~\cite{esser2010general}. LPD is an end-to-end deep unrolled method, capable of solving the linear inverse problem like (3) with the forward operator embedded in the network as shown in Algorithm \ref{alg1}. $\Gamma _{{\theta _d}}$ and $\Lambda _{{\theta _p}}$ are two convolutional neural networks with parameter ${\theta _d}$ and ${\theta _p}$. $A$ denotes forward projection and it's adjoint operator $A^*$ denotes the backward projection. LPD has achieved good results on many image to image reconstruction problem like CT, CS-MRI~\cite{wang2019accelerating} and PET~\cite{guazzo2021learned}. 

\begin{algorithm}
\caption{Learned primal dual (LPD).}
\label{alg1} 
\renewcommand{\algorithmicrequire}{\textbf{Input:}}
    \begin{algorithmic}[1] %这个1 表示从第一行开始显示行号，不写就不会显示行号
    \REQUIRE ~~Image initialization ${f_0}=\textbf{0}$, dual variable initialization ${h_0}=\textbf{0}$
    \FOR{$k \in [1,...,K]$}
    \STATE $h_k^{} \leftarrow {\Gamma _{{\theta _d}}}(h_{k - 1}^{},A{f_{k - 1}},g)$
    \STATE $f_k^{} \leftarrow {\Lambda _{{\theta _p}} }({f_{k - 1}},{A^*}{h_k})$
    \ENDFOR
    \RETURN $f^K$;
    \end{algorithmic}
\end{algorithm}

However, the deep unrolled methods, represented by LPD, can hardly be used in the TOF PET list-mode data. The main reasons are as follows: 1) For sinogram data, the dimensions of the system matrix are usually determined in advance, and the system matrix is typically precalculated and stored for use in the deep unrolled methods. In contrast, the dimensions of the system matrix for list-mode data are influenced by the counting rates, making it hard to precalculate and store the system matrix in advance, which affects the network training. 2) List-mode data records the location, energy and time information of coincidence events in the form of a binary list. While the convolutional operator in deep unrolled methods can effectively process image data such as sinograms, it cannot directly handle such binary list.

\begin{figure*}         
\centering
\includegraphics[width=\linewidth]{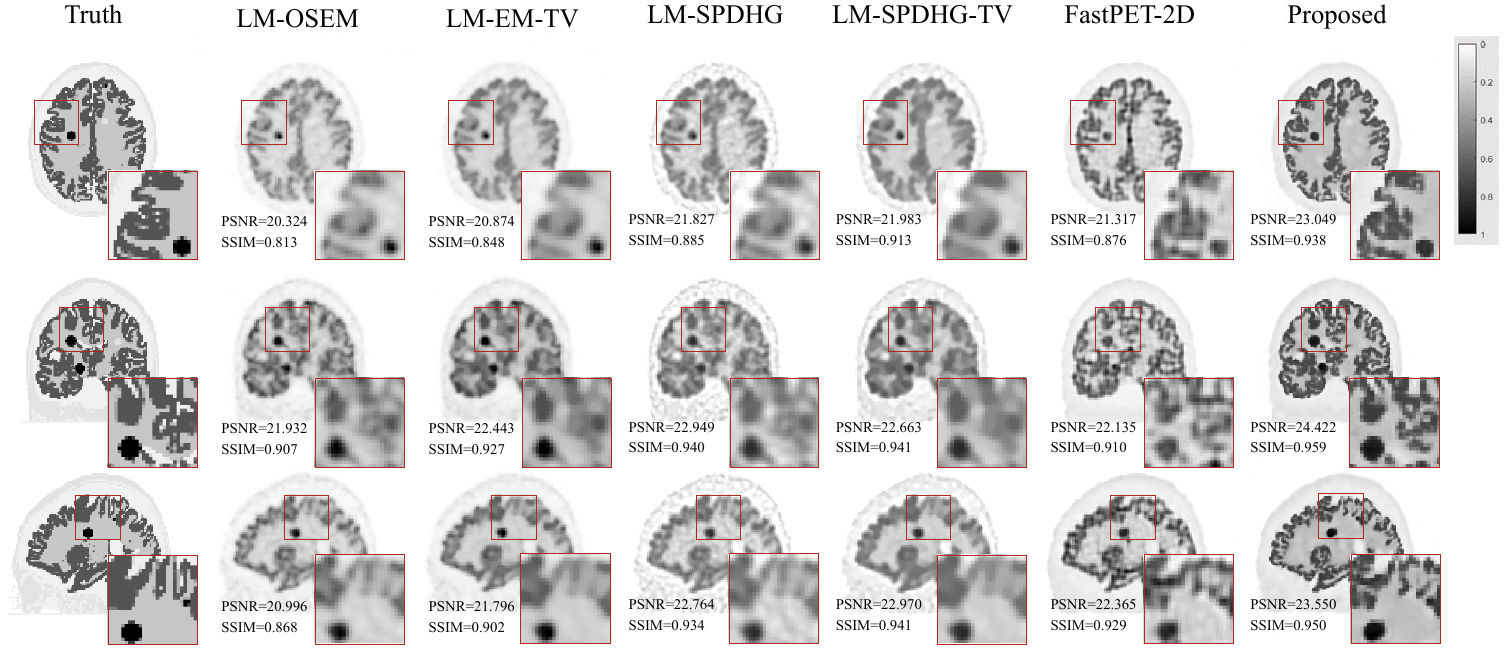}  
\caption{The ground truth images and the reconstructed images with 3e5 counts and TOF resolution of 200 ps by different reconstruction methods. From left to right: Truth, LM-OSEM, LM-EM-TV, LM-SPDHG, LM-SPDHG-TV, FastPET-2D and proposed LMPDnet.} 
\label{main_result}
\end{figure*}

\subsection{Proposed method}

Inspired by LPD, we propose an unrolled network for TOF list-mode reconstruction. The reconstruction scheme is shown in Fig.\ref{Fig1}. The entire reconstruction process are unrolled into several phases, with the input of each phase consisting of the list-mode domain variable $h_k$, the image domain variable $f_k$ and the measurement list-mode data $g$. The output of each phase includes updated $h_{k+1}$ and $f_{k+1}$, which are then fed into the next phase. 

Each phase contains a dual module for list-mode domain learning and a primal module for image domain learning. In dual module, the image domain variable $f$ is first forward-projected to the list-mode domain using TOF list-mode projection operator $Af_{k-1}$. The projected $f$ is then concatenated with the measurement list-mode data $g$ and the list-mode domain variable $h$. The concatenated tensor, comprising $Af_{k-1}$, $g$ and $h$, has a shape of $[N,3]$. This tensor is processed by a fully connected neural network, which in this work consists of three linear layers with PReLU activation.

The output of dual module (the updated list-mode domain variable $h$) is back-projected to the image domain using TOF list-mode back-projection operator. This output is then concatenated with the image domain variable $f$ and fed into a CNN. The CNN used in this study includes five convolutional layers, PReLU activation and batch normalization.

Unlike LPD, the proposed LMPDnet receives list-mode data rather than image as input. Given the sequence of coincidence events in list-mode data is random and disordered, and the length is flexible in each measurement, the TOF list-mode projection operator needs to be calculated on-the-fly. Therefore, we modified the TOF list-mode projection operator to enable CUDA-accelerated real-time computing. 
%合并了同一个TOF-bin的数据，减少显存占用
\subsection{TOF list-mode forward/backward projection}
For TOF list-mode data, the projection result of an event in $i$-th TOF bin is the integral of the activity on this TOF bin:
\begin{equation}
\label{equ3}
    h(i) = \int_{bin_{i}}\varepsilon_{i}\cdot f(p,q)ds
\end{equation}
where $f(\cdot)$ is the activity distribution, the $\varepsilon_{i}$ is the TOF weight and $p\in[1,P], q\in[1,Q]$ represents image coordinates. $ds$ represents the differential element of the TOF bin $i$ corresponding to this line integral. We model the TOF weights using a Gaussian distribution, with the expected annihilation location detected represented as $\mu _{TOF}$ and the standard deviation represented as $\sigma _{TOF}$, the $\varepsilon_{i}$ is as follows:
\begin{equation}
    {\varepsilon _i} = \frac{1}{2}erf(\frac{{{d_{TOF}} + \frac{1}{2}{\omega _{TOF}}}}{{\sqrt 2 {\sigma _{TOF}}}}) - \frac{1}{2}erf(\frac{{{d_{TOF}} - \frac{1}{2}{\omega _{TOF}}}}{{\sqrt 2 {\sigma _{TOF}}}})
\end{equation}
where $d_{TOF}$ denotes the distance from the midpoint of the TOF bin to $\mu _{TOF}$, and $\omega _{TOF}$ represents the width of the TOF bin. The error function $erf()$ is calculated as follows:
\begin{equation}
    erf(x) = \frac{x}{{|x|}}\sqrt {1 - \exp ( - {x^2}\frac{{4/\pi  + 0.14{x^2}}}{{1 + 0.14{x^2}}})}    
\end{equation}

\begin{figure}[]
\centering  
\includegraphics[width=\linewidth]{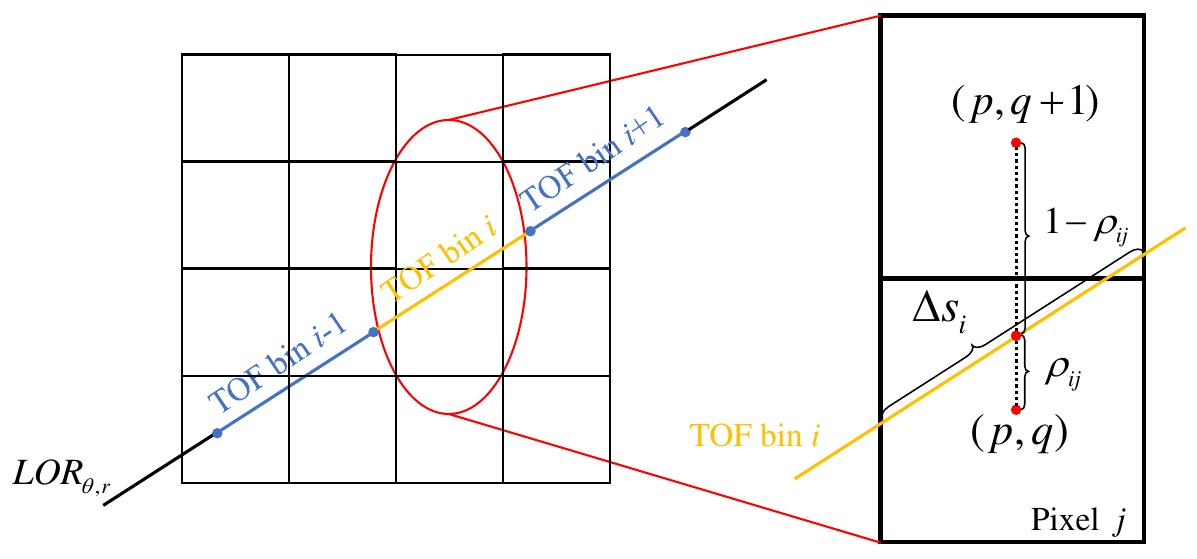}
\caption{The illustration of the calculation of TOF PET list-mode projection matrix.}
\label{figproj}
\end{figure}

The value of each element in the projection matrix can be obtained in the process of calculating (\ref{equ3}). The contribution of each pixel to the LOR varies linearly with the position, so the result of the line integration can be expressed as the product of the length of the line segment $\Delta s_i$ and the midpoint activity value of the line segment $f(s_{mid})$:

\begin{equation}
\label{eqhi}
    h(i) = \sum_{q=1}^{Q}\varepsilon_{i}\cdot f(s_{mid}) \cdot\Delta s_{i}
\end{equation}
The midpoint activity value $f(s_{mid})$ can be calculated by linear interpolation between two adjacent pixel, for example $(p, q)$ and $(p, q+1)$:
\begin{equation}
\label{equ5}
    f(s_{mid}) = \rho_{ij} \cdot f(p, q) + (1-\rho_{ij}) \cdot f(p, q + 1)
\end{equation}
where $\rho_{ij}$ is the coefficient of linear interpolation as shown in Fig. \ref{figproj}. Here we assume the pixel size is 1. By substituting this formula into Eq. \ref{eqhi}, we obtain:
\begin{equation}
\label{eq9}
h(i)=\sum_{p=1}^{P} \varepsilon_i \left(\rho_{ij} f \left(p,q \right) + (1-\rho_{ij})\cdot f\left(p,q + 1\right) \right) \Delta s_{i}   
\end{equation}
Considering that the projection process of the $i$-th TOF bin can also be obtained by multiplying the 
$i$-th row $A_i$ of the projection matrix with the image $f$:
\begin{equation}
\label{eq10}
    h(i)=\sum_{j=1} ^M a_{ij}f_j 
\end{equation}
Therefore, using Eq. \ref{eq9} and Eq. \ref{eq10}, the element values in the projection matrix are calculated as follows:
\begin{equation}
\label{equ11}
\begin{aligned}
A[i][q\cdot P+p] & =\varepsilon_i \cdot \rho_{ij} \cdot \Delta s_{i} \\
A[i][(q+1)\cdot P +p] & =\varepsilon_i \cdot(1-\rho_{ij}) \cdot \Delta s_{i}
\end{aligned}
\end{equation}
where $M = P \cdot Q$ and $j = q\cdot P + p$. 

In view of that there is no data interaction between the calculations of each LOR projections, so the granularity of each LOR projection calculation is relatively fine. We made the calculation of projection accelerated by CUDA parallel computing.

\subsection{Implementation details and reference methods}
The LMPDnet was implemented using Pytorch 1.6 on a NVIDIA GTX TITAN X (24 GB). The Adam optimizer with default settings was used to optimize the Mean Square Error (MSE) loss function between network output images and labels during training. The learning rate was $10^{-6}$. Due to the limitation of video memory size, the batch size was set to 1. The number of unrolled phases was 8 . For the dual module, two linear hidden layers with 64 and 16 features were used, and the primal module included 5 convolutional layers along with Batch Normalization and PReLU activation functions. The kernel size of each convolutional layer was 3$\times$3. The channel numbers following an encoder-decoder design, are 2, 64, 128, 256, 64, and 1. Based on the Joseph projection calculation method~\cite{schramm2022parallelproj,joseph1982improved}, CUDA was used to compute the projection of each line of response (LOR) of the incoming list-mode event in parallel to achieve acceleration. The total number of training epochs were 500, and the trained model with the minimum validation loss was chosen as the final reconstruction model. The proposed LMPDnet was compared with list-mode ordered subset expectation maximization (LM-OSEM), total variation regularized EM (LM-EM-TV),  list-mode primal-dual hybrid gradient algorithm (LM-PDHG)~\cite{schramm2022fast}, total variation regularized PDHG (LM-SPDHG-TV) and histogram based TOF list-mode image reconstruction method FastPET~\cite{whiteley2020fastpet}. The subsets were set to 4 for LM-OSEM and LM-EM-TV and set to 224 for LM-PDHG and LM-PDHG-TV. For LM-OSEM and LM-EM-TV, the iteration number is 15 and the algorithm converged. For LM-SPDHG and LM-SPDHG-TV, the iteration number is 5. The TV regularization parameter $\beta$ was set to 2.0 for LM-EM-TV and 0.20 for LM-SPDHG-TV. The selection of hyper-parameters is shown in Fig.\ref{hyper}. The step size ratio $\gamma$ was adjusted in relation to the count number and $\rho_{ij}$ was set to 0.999. Given that our current experiments are limited to 2D scenarios, we adapted FastPET to a 2D version. The learning rate for FastPET-2D was $10^{-3}$, batch size is 32 and total traning epoch is 2000. 

\begin{figure*}[]
\centering  
\subfigure[]{
\label{Fig.sub.1}
\includegraphics[width=7cm]{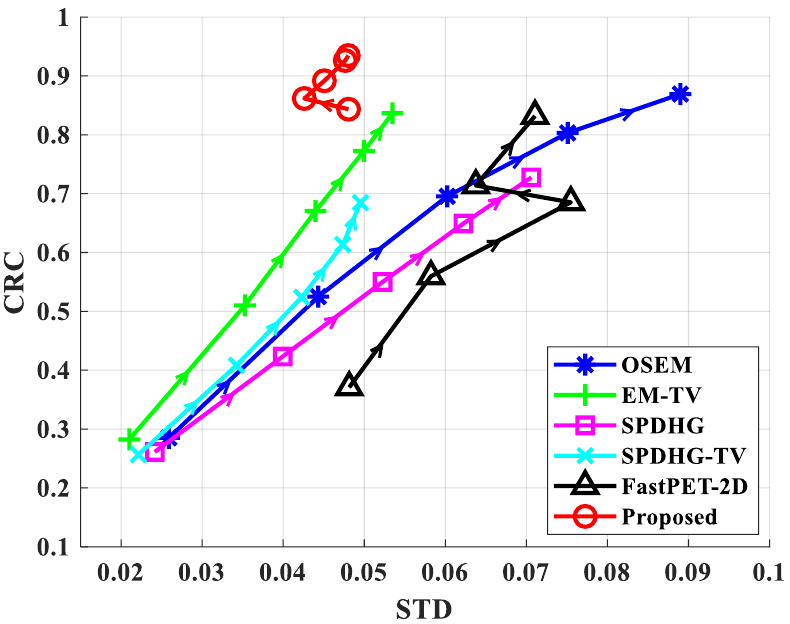}}\subfigure[]{
\label{Fig.sub.2}
\includegraphics[width=7cm]{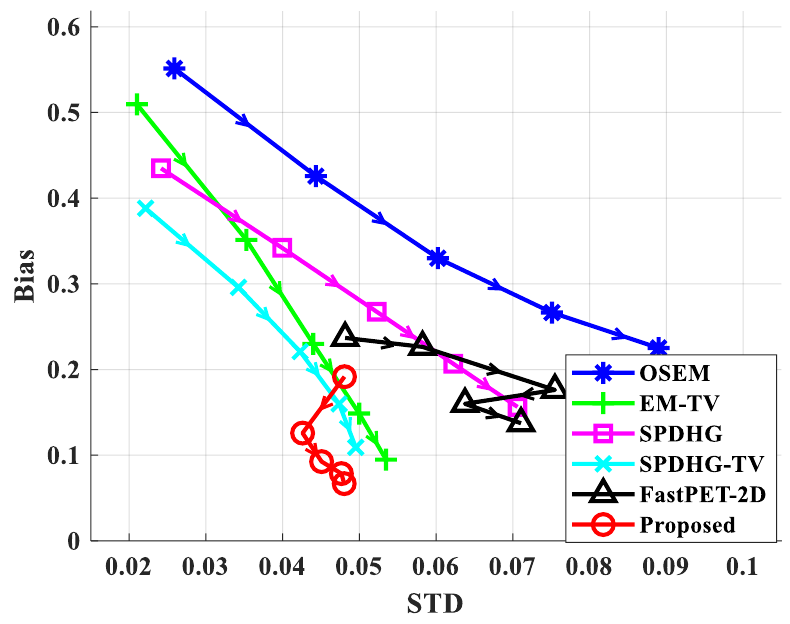}}
\caption{The CRC-STD and Bias-STD trade-off curves between contrast and noise in the simulation study for different methods. (a) CRC vs. STD curve; (b)Bias vs. STD curve. For OSEM and EM-TV, markers were plotted every 3 iterations. For SPDHG and SPDHG-TV, markers were plotted every 1 iterations. For FastPET-2D, markers were plotted every 400 epochs and for proposed LMPDnet, markers were plotted every 2 phases.}
\label{Fig5}
\end{figure*}

\section{Experiments and results} 
\subsection{Experimental setup}
\subsubsection{Brain phantom simulation}
We employed 20 3D brain phantoms from BrainWeb~\cite{cocosco1997brainweb} to simulate 2D ${}^{18}$F-FDG PET images. Following the simulation settings in the previous paper~\cite{mehranian2020model}, We segmented the T1-weighted MR images from BrainWeb into gray matter (GM), white matter (WM), cerebrospinal fluid (CSF), skull and skin. For each image, the FDG PET phantom was generated as follows: random uptake values of $96.0 \pm 5.0$ and $32.0 \pm 5.0$ were assigned to the GM and WM regions, respectively. Fifteen spherical hot regions with radius ranging from 2mm to 8mm were embedded in all of the phantoms. An uptake value of 144 (1.5 $\times$ of GM) was assigned to the hot lesions, and an uptake value of 48.0 (0.5 $\times$ of WM) was assigned to the cold lesions. The PET images were generated with a resolution 2.086$\times$2.086$\times$2.031 $mm^3$ and a matrix size of 334$\times$334$\times$127. Twenty non-continuous slices were selected from each of the three orthogonal views for each phantom to generate the TOF list-mode data with different counts level (1e5, 3e5 and 1e6), different TOF resolution (200 ps, 300 ps and 400 ps) and different number of the TOF bins (5,11,17). To adequately cover the object, with a TOF bin number of 5, the bin size is 51mm; with a number of 11, the bin size is 23.2mm; and with a number of 17, the bin size is 15mm. Each 2D images were resized to 128$\times$128. We incorporated the effects of limited spatial resolution and attenuation, as well as a flat contamination that mimics both random and scattered coincidences, with a contamination fraction of 20\%. The PET raw data used in this study were acquired using a cylindrical scanner consisting of 28 modules, each containing 16 crystals with a radial width of 4 mm. As a result, the sinogram dimension was 357$\times$224$\times$17. Noisy simulated prompt emission TOF sinograms and the corresponding list-mode data were generated, making a total of 1200 list-mode image pairs for each count level and TOF resolution. Among them, 17 brain samples (1020 list-mode image pairs) were selected as training data, 2 (120 list-mode image pairs) as testing data and 1 (60 list-mode image pairs) as validation data.

The peak signal-to-noise ratio (PSNR) and structural similarity index (SSIM) were employed to assess overall image quality. For quantitative comparison and bias-variance analysis, contrast recovery coefficient (CRC)~\cite{qi1999theoretical} versus standard deviation (STD) curves, as well as bias versus standard deviation curves, were plotted.

\begin{table}[]
\caption{The quantitative analysis of different methods on different count level with the TOF resolution of 200 ps and 17 TOF bins}
\label{diff_count}
\resizebox{\linewidth}{!}{%
\begin{tabular}{@{}ccccccc@{}}
\toprule[1.5pt]
\multicolumn{1}{c}{\multirow{2}{*}{Method}} & \multicolumn{2}{c}{Count = 1e6} & \multicolumn{2}{c}{Count = 3e5} & \multicolumn{2}{c}{Count = 1e5} \\ \cmidrule(l){2-7} 
\multicolumn{1}{c}{} & PSNR       & SSIM        & PSNR       & SSIM        & PSNR       & SSIM        \\ \cmidrule(r){1-7}
LM-OSEM                 & 20.55±2.01 & 0.877±0.037 & 20.08±2.25 & 0.855±0.050 & 19.17±2.48 & 0.8070±0.073 \\
LM-EMTV                & 20.93±1.88 & 0.900±0.027 & 20.56±2.06 & 0.886±0.035 & 19.91±2.39 & 0.857±0.056 \\
LM-SPDHG                & 21.80±1.92 & 0.922±0.024 & 21.21±2.08 & 0.902±0.028 & 19.39±1.67 & 0.881±0.027 \\
LM-SPDHG-TV             & 21.86±2.01 & 0.928±0.021 & 21.24±2.06 & 0.911±0.027 & 19.89±1.60 & 0.897±0.039 \\
FastPET-2D             & 21.36±3.17 & 0.890±0.085 & 20.72±2.96 & 0.876±0.086 & 19.94±2.71 & 0.856±0.082 \\
Proposed             & \textbf{24.30}±2.45 & \textbf{0.957}±0.023 & \textbf{22.22}±2.15 & \textbf{0.933}±0.025 & \textbf{20.66}±1.90 & \textbf{0.907}±0.029 \\
\bottomrule[1.5pt]
\end{tabular}%
}
\end{table}

\begin{table}[]
\caption{The quantitative analysis of different methods on different TOF resolution with the count of 3e5 and 17 TOF bins}
\label{diff_TOFreso}
\resizebox{\linewidth}{!}{%
\begin{tabular}{@{}ccccccc@{}}
\toprule[1.5pt]
\multicolumn{1}{c}{\multirow{2}{*}{Method}} & \multicolumn{2}{c}{200 ps} & \multicolumn{2}{c}{300 ps} & \multicolumn{2}{c}{400 ps} \\ \cmidrule(l){2-7} 
\multicolumn{1}{c}{} & PSNR       & SSIM        & PSNR       & SSIM        & PSNR       & SSIM        \\ \cmidrule(r){1-7}
LM-OSEM                 & 20.07±2.25 & 0.855±0.050 & 20.03±2.10 & 0.851±0.0383 & 19.98±2.10 & 0.848±0.035 \\
LM-EMTV                & 20.56±2.06 & 0.886±0.035 & 20.50±1.91 & 0.884±0.025 & 20.39±1.84 & 0.877±0.022 \\
LM-SPDHG                & 21.21±2.08 & 0.902±0.028 & 20.99±1.96 & 0.898±0.022 & 20.78±1.87 & 0.879±0.020 \\
LM-SPDHG-TV             & 21.24±2.06 & 0.911±0.027 & 21.13±1.95 & 0.911±0.022 & 21.06±1.83 & 0.903±0.019 \\
FastPET-2D             & 20.72±2.96 & 0.876±0.086 & 20.62±3.06 & 0.864±0.102 & 20.29±2.80 & 0.864±0.088 \\
Proposed             & \textbf{22.22}±2.15 & \textbf{0.933}±0.025 & \textbf{21.62}±2.07 & \textbf{0.923}±0.027 & \textbf{21.29}±1.88 & \textbf{0.906}±0.031 \\
\bottomrule[1.5pt]
\end{tabular}%
}
\end{table}

\begin{table}[]
\caption{The quantitative analysis of different methods on different number of TOF bins with the TOF resolution of 200 ps and the count level of 3e5.}
\label{diff_TOFbin}
\resizebox{\linewidth}{!}{%
\begin{tabular}{@{}ccccccc@{}}
\toprule[1.5pt]
\multicolumn{1}{c}{\multirow{2}{*}{Method}} & \multicolumn{2}{c}{17 TOF bins} & \multicolumn{2}{c}{11 TOF bins} & \multicolumn{2}{c}{5 TOF bins} \\ \cmidrule(l){2-7} 
\multicolumn{1}{c}{} & PSNR       & SSIM        & PSNR       & SSIM        & PSNR       & SSIM        \\ \cmidrule(r){1-7}
LM-OSEM                 &20.08±2.25 & 0.855±0.050 & 19.89±2.17 & 0.861±0.044 & 19.82±2.18 & 0.872±0.042 \\
LM-EMTV                & 20.56±2.06 & 0.886±0.035 & 20.21±2.02 & 0.890±0.031 & 20.08±1.97 & 0.894±0.028 \\
LM-SPDHG                & 21.22±2.08 & 0.902±0.030 & 21.16±2.08 & 0.910±0.027 & 21.03±2.00 & 0.901±0.025 \\
LM-SPDHG-TV             & 21.24±2.06 & 0.911±0.027 & 21.19±1.89 & 0.917±0.022 & 21.09±1.78 & 0.911±0.023 \\
FastPET-2D             & 20.72±2.96 & 0.876±0.086 & 20.63±2.43 & 0.869±0.075 & 20.32±2.82 & 0.862±0.120 \\
Proposed             & \textbf{22.22}±2.15 & \textbf{0.933}±0.025 & \textbf{21.90}±2.07 & \textbf{0.928}±0.025 & \textbf{21.35}±1.98 & \textbf{0.918}±0.027 \\
\bottomrule[1.5pt]
\end{tabular}%
}
\end{table}

\begin{figure*}
\centering
\includegraphics[width=\linewidth]{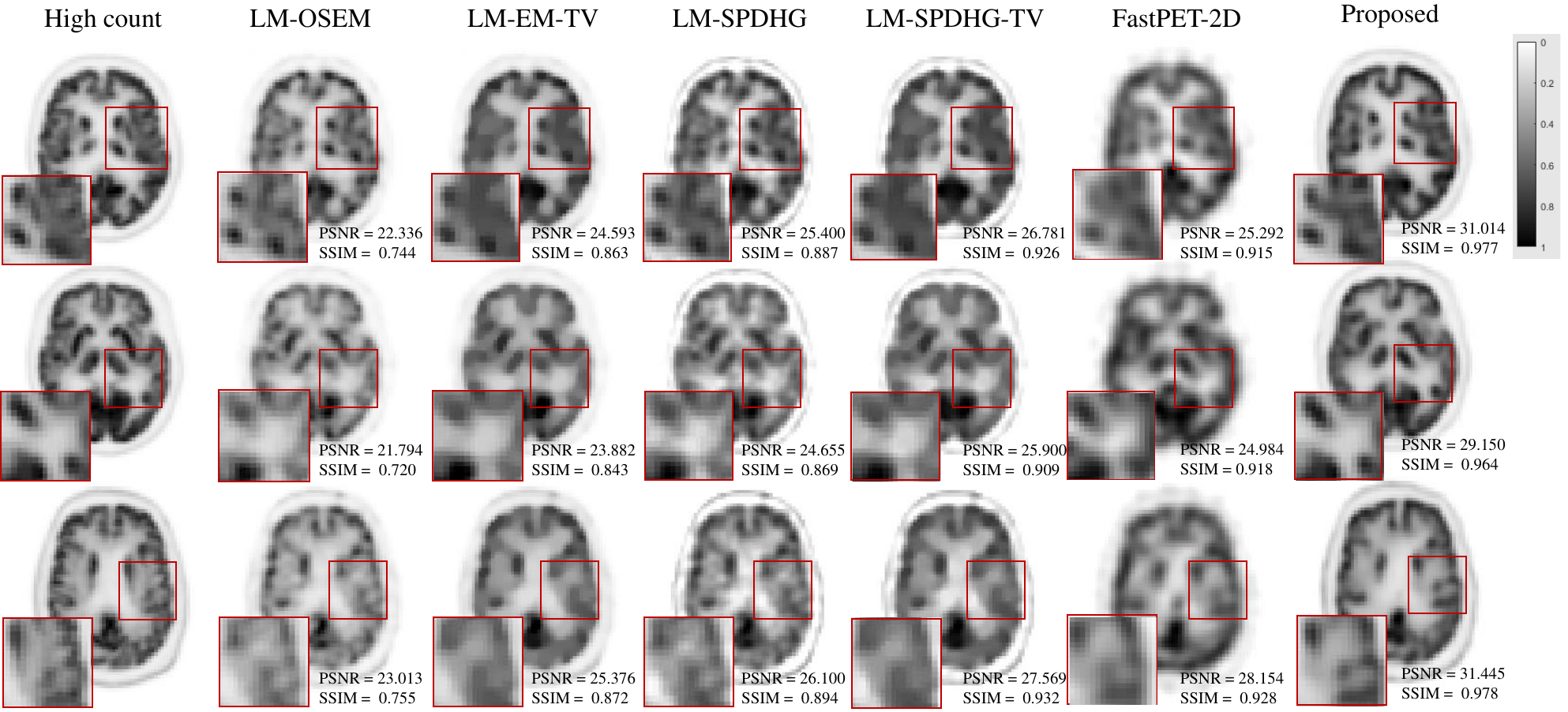}  
\caption{The High count clinical images reconstructed by EM and the reconstructed images with 3e5 counts and TOF resolution of 200 ps by different reconstruction methods. From left to right: High count EM, OSEM, EM-TV, SPDHG, SPDHG-TV, FastPET-2D and proposed LMPDnet.} 
\label{clinical}
\end{figure*}

The PSNR can be expressed as:
\begin{equation}
    M S E=\sqrt{\frac{1}{N} \sum_{i=1}^N\left(A_i-B_i\right)^2}
\end{equation}

\begin{equation}
    P S N R=20 \cdot \log _{10}\left(\frac{B_{\max }}{M S E}\right)
\end{equation}

In this equation, $A$ represents the reconstructed image, $B$ denotes the ground truth, $N$ stands for the number of image pixels, and $B_{\max}$ is the maximum value of the ground truth image.

The SSIM is formulated as:
\begin{equation}
S S I M=\frac{\left(2 \mu_A \mu_B+c_1\right)\left(2 \sigma_{A B}+c_2\right)}{\left(\mu_A^2+\mu_B^2+c_1\right)\left(\sigma_A^2+\sigma_B^2+c_2\right)}
\end{equation}
Here, $\mu$ and $\sigma$ signify the mean and variance of the image, respectively. The constants are defined as $C_1 = 0.01 \times \max (A)$ and $C_2 = 0.03 \times \max (A)$.

The CRC is given by:
\begin{equation}
C R C=\frac{1}{S} \sum_{s=1}^S \frac{\left(\frac{\bar{a}_s}{\bar{b}_s}-1\right)}{\left(\frac{a_{\text {true }}}{b_{\text {true }}}-1\right)}
\end{equation}
In this formula, $S$ represents the number of realizations, $\bar{a}s$ and $\bar{b}_s$ are the average values of selected regions of interest (ROIs) and background regions in the $s$-th realization, and $a_{true}$ and $b_{true}$ are the ground truth values for the target and background regions, respectively. In the simulation study, the number of realizations is set to 5. Two tumor regions were selected as target ROIs, while eight four-pixel-diameter spheres were drawn in white matter as background regions. The background STD is computed as:
\begin{equation}
S T D=\frac{1}{K_b} \sum_{k=1}^{K_b} \frac{\sqrt{\frac{1}{S-1} \sum_{s=1}^S\left(b_{s, k}-\bar{b}_s\right)^2}}{\bar{b}_k}
\end{equation}
In this equation, $K_b = 15$ denotes the total number of ROIs in the background region, and $\bar{b}_k=$ $(1 / S) \sum_{s=1}^S b_{s, k}$ represents the average of the $k$-th ROI over $S$ realizations.

The bias of the mean values in the entire target region is defined as:
\begin{equation}
Bias=\frac{\bar{A}-B}{A}
\end{equation}
Here, $\bar{A}$ is the average value of the entire target region for all realizations, and $B$ signifies the ground truth value of the whole target region.

\subsubsection{Clinical image simulation}
The $^{18}$F-FDG total-body datasets of forty patients acquired from Siemens Biograph Vision Quadra were used in this study\cite{UltraLowDosePET2023}.  Thirty brain samples were randomly chosen for training, five for testing and another five for validation. Additionally, five thorax samples were selected for validating generalization. Each brain sample contains forty image slices. All the images were 128 pixels $\times$ 128 pixels and normalized using the global min-max scaling. We forward-projected and down-sampled the data to 3e5 counts with 200 ps TOF resolution and 17 TOF bins to generate low-count situations.
During quantitative evaluation, the contrast to noise ratio (CNR) was employed and defined as:
\begin{equation}
    CNR =\frac{ (mean_{ROI} - mean_{background})}{STD_{background}}
\end{equation}
where $mean_{background}$ denotes the mean intensity in the background regions, $STD_{background}$ denotes the standard deviation in the background regions. $mean_{ROI}$ denotes the mean intensity in the ROI region. In patient brain data, the gray matter was selected as the ROI region and the white matter was selected as the background region.

\begin{figure}[]
\centering  
\subfigure[]{
\label{CNR.sub.1}
\includegraphics[width=\linewidth]{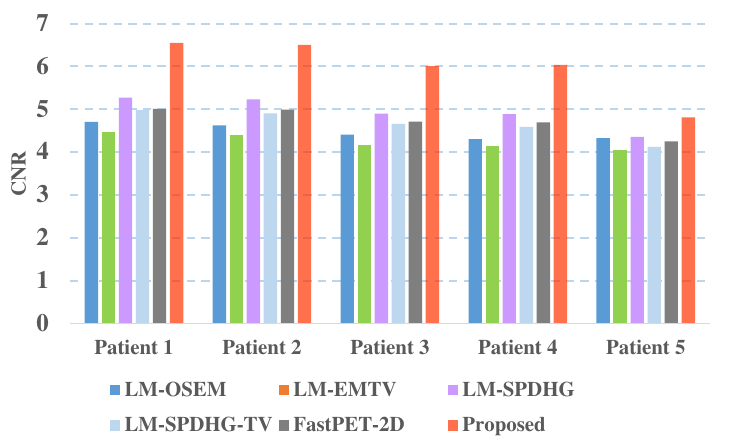}} \\ \subfigure[]{
\label{CNR.sub.2}
\includegraphics[width=\linewidth]{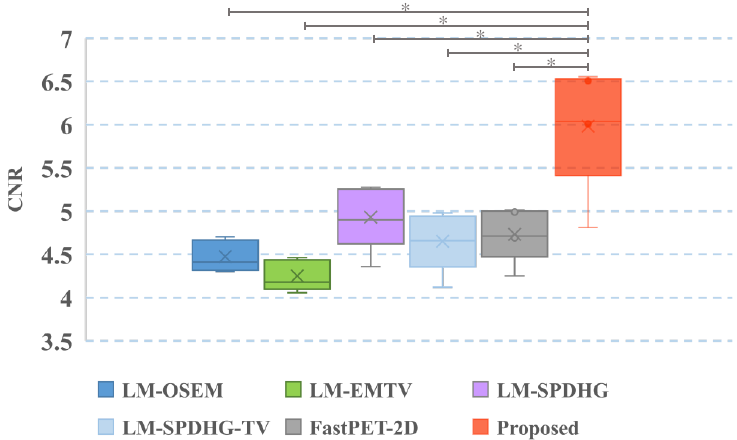}}
\caption{The contrast to noise ratio (CNR) for 5 patients' test data. (a) The CNR of different methods in different patient. (b) The boxplot of the CNR for all 5 patients of different methods. In the boxplots, * representing P $<$ 0.05.  }
\label{CNR}
\end{figure}

\subsection{Results}

% \begin{figure}         
% \centering
% \includegraphics[width=\linewidth]{figures/diffTOF_result.pdf}  
% \caption{Image reconstructed by different methods with different TOF resolution (400ps, 800ps and 1200ps). From left to right, Ground truth, OSEM, EM-TV, SPDHG, SPDHG-TV and Proposed LMPDnet.} 
% \label{Fig3}
% \end{figure}
% 
Fig. \ref{main_result} shows the images reconstructed by different methods. The reconstruction results of LM-OSEM exhibit noticeable noise, and the structural information is quite blurred. LM-EMTV effectively removes noise but exhibits apparent over-smoothing, leading to the loss of reconstructed cortical edge structures. Compared to the EM-based results, LM-SPDHG reconstruction yields more accurate activity map estimates; however, it also presents significant artifacts, particularly at contour edges. LM-SPDHG-TV alleviates the severe noise issue in LM-SPDHG but similarly displays over-smoothing. FastPET-2D slightly outperforms the traditional methods in terms of detailed structure recovery, but the overall image still exhibits noticeable noise. The proposed LMPDnet achieves the best performance in terms of noise suppression and structural information recovery.

Tab. \ref{diff_count} shows the quantitative analysis of the images reconstructed by different methods with different count data. The impact on image quality due to a reduction in count rate is pronounced. Under low count rate conditions, the PSNR of LM-OSEM and SPDHG is relatively low. EM-TV and SPDHG-TV slightly raise the PSNR with regularization. The proposed LMPDnet displays excellent noise robustness, showing great potential for application in low-count reconstruction.

Tab. \ref{diff_TOFreso} shows the quantitative analysis of the images reconstructed by different methods with different TOF resolution. As the TOF time resolution decreases, both LM-OSEM and LM-SPDHG reconstruction results exhibit a certain degree of structural blurring, and the PSNR and SSIM decrease. In contrast, the proposed LMPDnet consistently achieves the closest performance to the true values across various TOF resolutions.

Tab. \ref{diff_TOFbin} shows the quantitative analysis of different methods on data with different numbers of TOF bins. In accordance with the reduction in the number of TOF bins, each method exhibits a decline in both PSNR and SSIM. The proposed LMPDnet achieves the best PSNR and SSIM performance as well as excellent robustness among the five methods.

Fig. \ref{Fig5} shows the CRC-STD and Bias-STD trade-off curves between contrast and noise for different methods. As seen in the figure, as the number of iterations increases, the tumor CRC of traditional methods rises; however, the standard deviation also exhibits a noticeable increase, the same effect is observed in the curve for FastPET-2D. Among the traditional methods, EM-TV achieves a relatively favorable CRC. With the increase in the number of phases for the proposed LMPDnet, it attains the best CRC performance, while the standard deviation remains similar value, demonstrating the stability of the reconstructed image. In terms of the overall image bias, SPDHG-TV achieves a commendable bias-variance performance among the traditional methods. As the number of iterations increases, the bias gradually decreases, while the standard deviation of traditional methods gradually rises. As the number of phases for the proposed LMPDnet increases, it exhibits the lowest bias, and the standard deviation remains very close.

Fig. \ref{clinical} presents the reconstruction results of different methods for low-count clinical data and the comparison with the results of the high-count EM. It can be observed that the proposed method yields better structural recovery and clearer details in high-uptake regions.s The results of quantitative analysis are presented in Table. \ref{tab2}, which shows that the proposed method achieved the best PSNR and SSIM performance.

Fig. \ref{CNR.sub.1} shows the CNR for all 5 patients' test data using different methods. The mean CNRs for LM-OSEM, LM-EMTV, LM-SPDHG, LM-SPDHG-TV, FastPET-2D and proposed LMPDnet are 4.474, 4.249, 4.930, 4.651, 4.732 and 5.983. Fig \ref{CNR.sub.2} shows the boxplots of the CNR of different methods, indicating that the CNR value of the proposed method is significantly higher than all comparison methods with P value less than 0.05.

\begin{table}[]
\caption{The quantitative analysis of different methods on clinical data.}
\label{tab2}
\begin{tabular}{@{}ccc@{}}
\toprule[1pt]
         & PSNR(dB)   & SSIM        \\ \midrule
LM-OSEM     & 23.83±2.01 & 0.942±0.016 \\
LM-EMTV    & 26.63±2.69 & 0.956±0.014 \\
LM-SPDHG    & 27.42±2.69 & 0.948±0.010 \\
LM-SPDHG-TV & 28.63±2.37 & 0.951±0.009 \\
Fast-PET-2D & 28.32±1.89 & 0.946±0.015 \\
Proposed & \textbf{32.17}±2.73 & \textbf{0.986}±0.004 \\ 
\bottomrule[1pt]
\end{tabular}%
\centering
\end{table}

\section{Discussion}

In this study, we proposed the first deep unrolled method for TOF PET list-mode reconstruction. The large system matrix and memory consumption caused by the unrolled method have always been significant challenges, especially in TOF reconstruction, where the additional space consumption due to TOF information makes unrolling ineffective. TOF information leads to almost exponential increases in sinogram data storage, but for list-mode data, processing TOF information is much easier. Table. \ref{computational} compares the storage space and computation time of the system matrix for list-mode and sinogram data in this study. Despite the advantages of list-mode data in TOF reconstruction, the system matrix computation for list-mode data and deep neural networks pose significant challenges for storage space. This study utilized a dynamic access method to calculate the system matrix for list-mode data on-the-fly, demonstrating the outstanding reconstruction performance of the deep unrolled method on TOF PET list-mode data.

\begin{figure}[]
\centering  
\subfigure[]{
\label{Fig.sub.1}
\includegraphics[width=8.4cm]{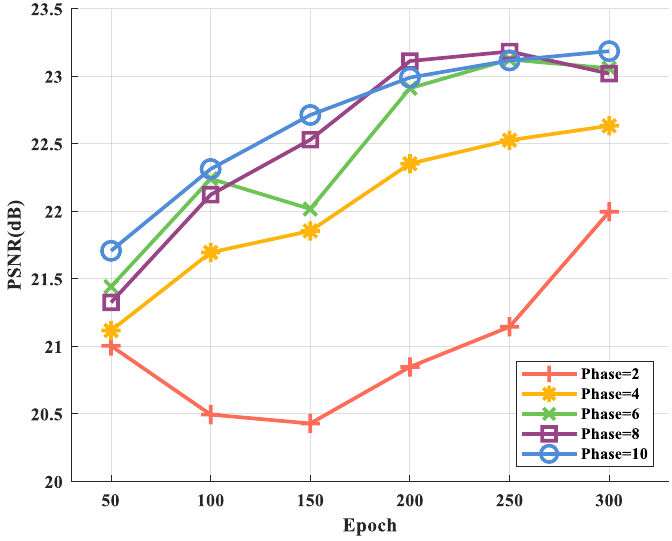}} 
\subfigure[]{
\label{Fig.sub.2}
\includegraphics[width=8.4cm]{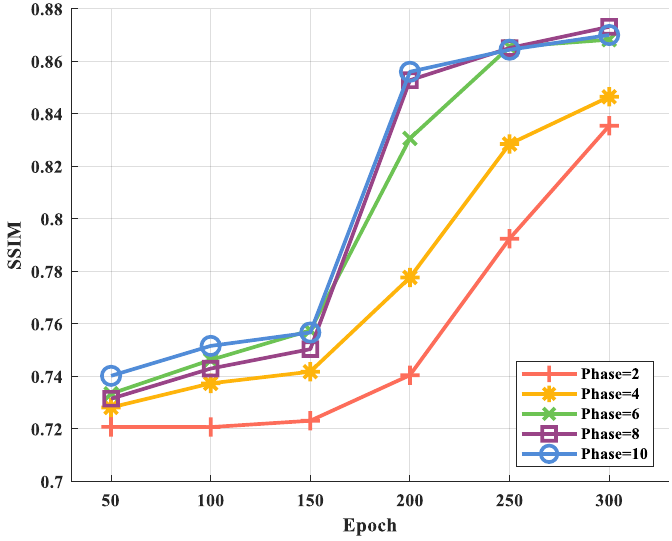}}
\caption{Ablation study on unrolled phase numbers, markers were plotted every 50 epochs. (a) The mean PSNR of test set with different phase numbers of LMPDnet. (b) The mean SSIM of test set with different phase numbers of LMPDnet.}
\label{differ_phase}
\end{figure}

\begin{table}[b]
\centering
    \caption{\label{table2}Comparison of video memory occupation and projection time consuming}
    \resizebox{8.5cm}{!}{
    \begin{tabular}{ccc}
    \toprule[1pt] %[2pt]
        &  sinogram & list-mode \\
    \midrule %[2pt]
        Memory Occupation & $\sim$87GB & $\sim$20GB \\
        Forward Projection & 364ms & 256ms \\
        Backward Projection & 405ms & 246ms \\
        Compute Projection Matrix & - & 157ms\\
    \bottomrule[1pt] %[2pt]
    \end{tabular}
    }
\label{computational}
\end{table}

In the design of the dual module, considering that both its input and output are list-mode data, we tried both Recurrent Neural Network (RNN) and Multi Layer Perceptron (MLP) as the backbone network for learning in the list-mode domain. We found that MLP outperformed RNN. The memory characteristics of RNN do not have practical significance between the different events collected in list-mode data as the each event is independent of the others.

The number of phases significantly impacts deep unfolding methods. Although a larger number of phases can provide greater learning ability, it also brings a heavy storage burden. We tested LMPDnet with different phase numbers, and Fig. \ref{differ_phase} shows the changes in PSNR and SSIM with the number of training epochs. When the phase number is small, both PSNR and SSIM significantly improve as the phase number increases. However, when the phase number is greater than or equal to 8, the final results are relatively close. Therefore, considering the trade-off between storage and effectiveness, we set the phase number to 8 in the experiment.

\begin{figure}         
\centering
\includegraphics[width=\linewidth]{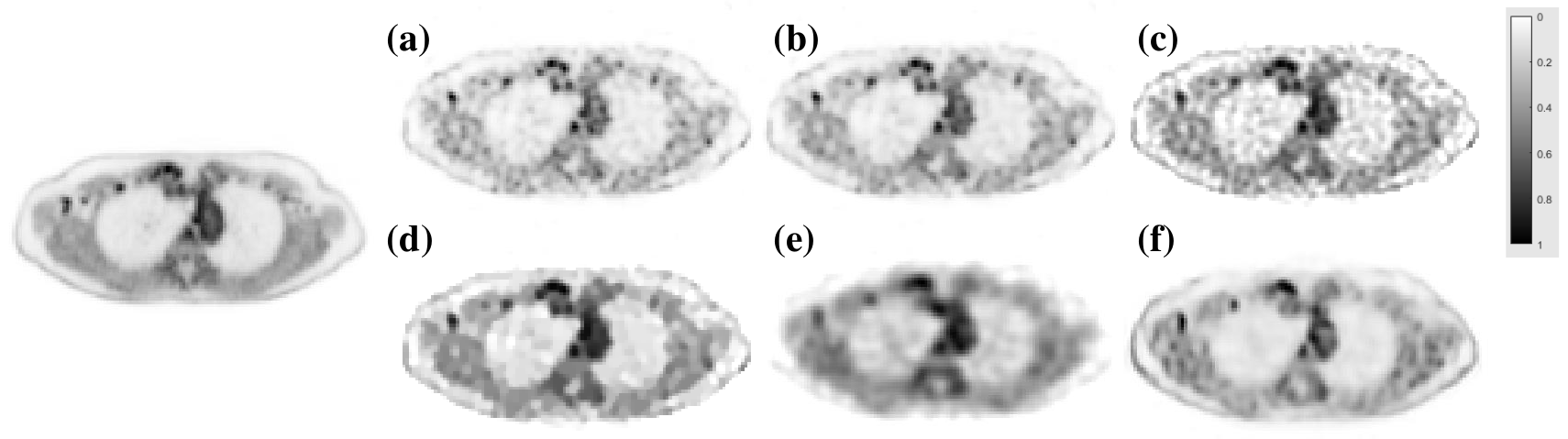}  
\caption{The reconstruction results of the thorax data by different methods. The count level is 3e5, the TOF resolution is 200ps and the number of TOF bin is 17. The image on the left is the hight count image. (a) LM-OSEM, (b) LM-EMTV, (c) LM-SPDHG, (d) LM-SPDHG-TV, (e)FastPET-2D, (f) Proposed. The proposed method and FastPET-2D are trained with only brain image data, and tested with thorax data.} 
\label{robust}
\end{figure}

\begin{figure}         
\centering
\includegraphics[width=\linewidth]{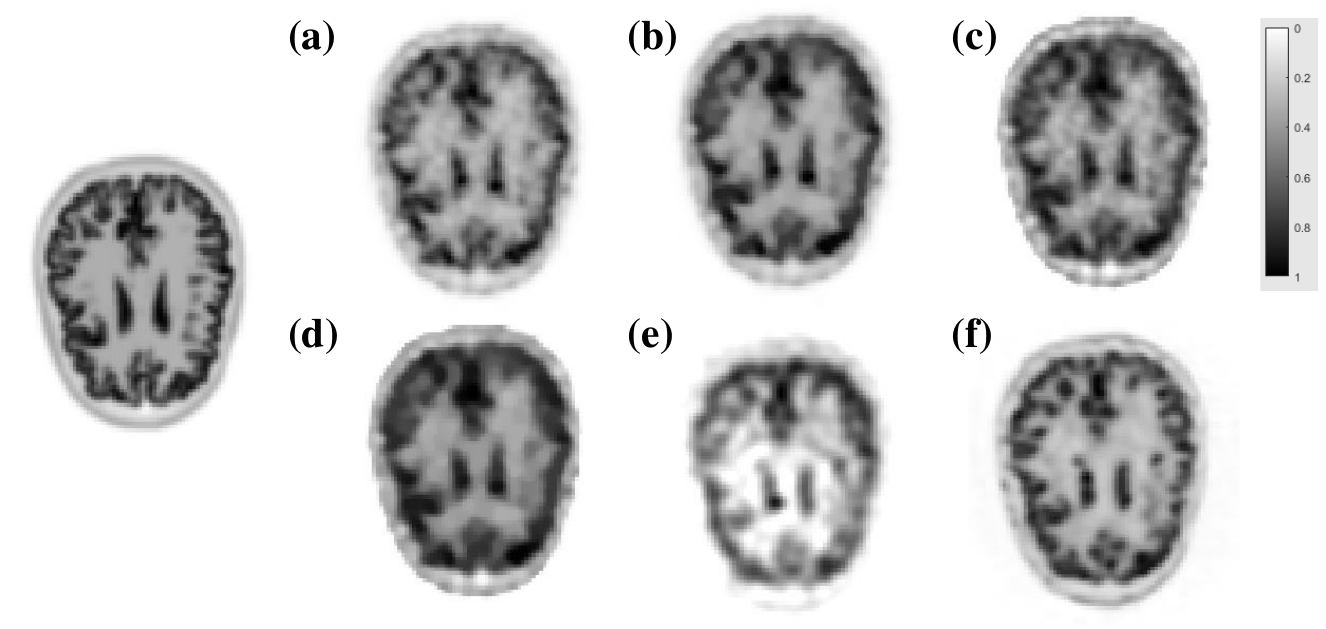}  
\caption{The reconstruction results of the C11-DASBSB image by different methods. (a) LM-OSEM, (b) LM-EMTV, (c) LM-SPDHG, (d) LM-SPDHG-TV, (e)FastPET-2D, (f) Proposed. The proposed method and FastPET-2D are trained with $^{18}$F-FDG image data, and tested with C11-DASBSB data.} 
\label{robust_tracer}
\end{figure}

Generalizability is a paramount consideration for the practical application of deep learning-based reconstruction algorithms. Deep unrolled methods have been demonstrated in numerous studies to possess superior generalization capabilities compared with end-to-end direct learning techniques. Fig. \ref{robust} shows the reconstruction results of different methods for thorax data. Notably, the proposed LMPDnet and FastPET-2D were trained exclusively using brain data, without any prior knowledge or exposure to thorax data. It is evident that even when trained solely on brain data, the proposed method exhibits commendable generalization performance on thorax data. Fig. \ref{robust_tracer} shows the reconstructions results of different methods for another tracer image (C11-DASBSB)~\cite{ds005138:1.0.2}. It can be seen that with the same small amount of training data, the generalization ability of the proposed LMPDnet greatly surpasses that of the FastPET-2D method.

\begin{table}[]
\caption{Computational cost of different methods (Unit: Second)}
\label{time}
\resizebox{\linewidth}{!}{%
\begin{tabular}{@{}ccccccc@{}}
\toprule[1.5pt]
Method & LM-OSEM & LM-EMTV & LM-SPDHG & LM-SPDHG-TV & FastPET-2D & Proposed \\ \midrule
Time   & 0.5974  & 0.8457  & 5.2040  & 6.1765     & 0.1804 & 0.4952   \\ 
\bottomrule[1.5pt]
\end{tabular}%
}
\end{table}

\begin{figure}         
\centering
\includegraphics[width=\linewidth]{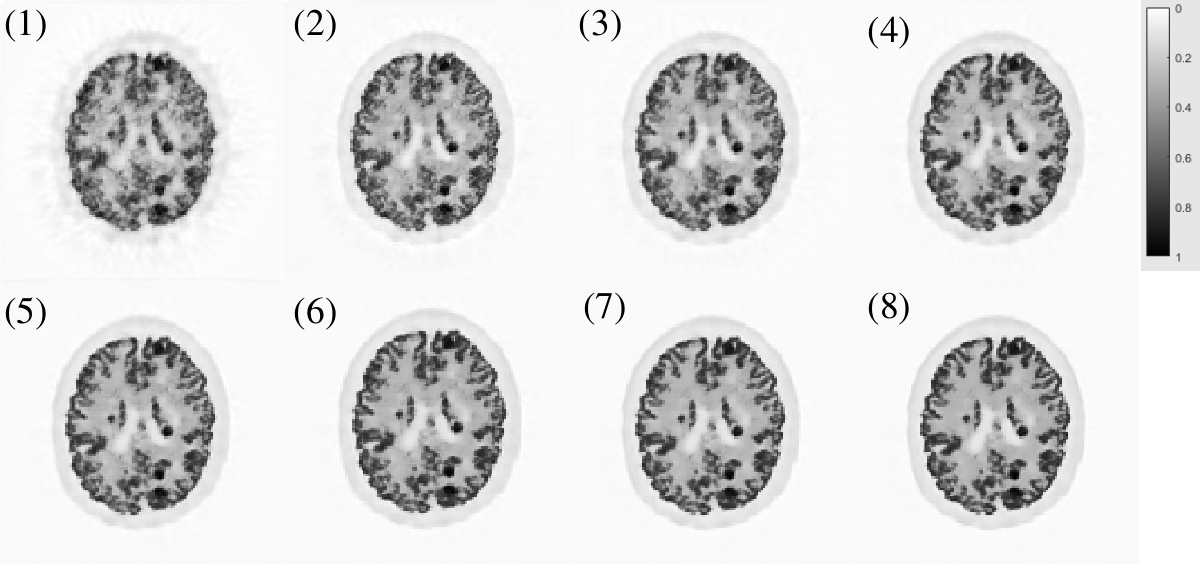}  
\caption{The output of each phases in the proposed method for simulation data, From (1) to (8): the output of the first phase to the output of the last phase.} 
\label{vis}
\end{figure}

Fig. \ref{vis} shows the output of each phase in the proposed method for the simulation data. In the initial few phases, the images display pronounced radial artifacts, especially in the first phase. As the number of phases increases, these artifacts progressively diminish, and the structural details of the image are gradually enhanced. Notably, the output of each phase adeptly captures the structural information of the image, further underscoring the robustness of the proposed method.

\begin{figure}[]
\centering  
\subfigure[]{
\label{Fig10.sub.1}
\includegraphics[width=4.2cm]{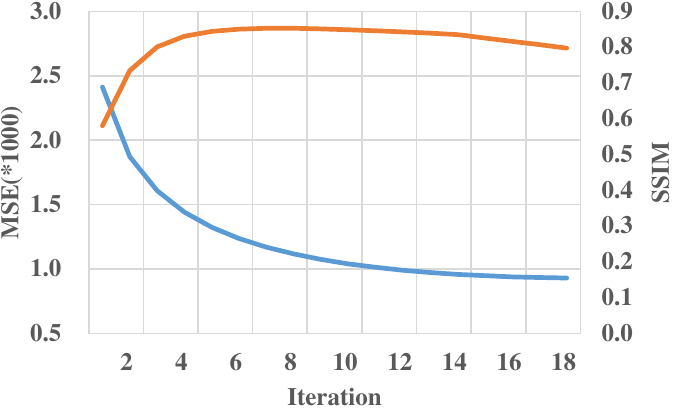}} 
\subfigure[]{
\label{Fig10.sub.2}
\includegraphics[width=4.2cm]{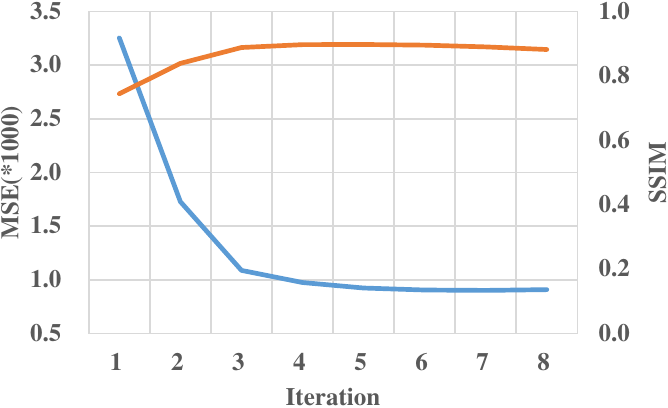}}
\subfigure[]{
\label{Fig10.sub.3}
\includegraphics[width=4.2cm]{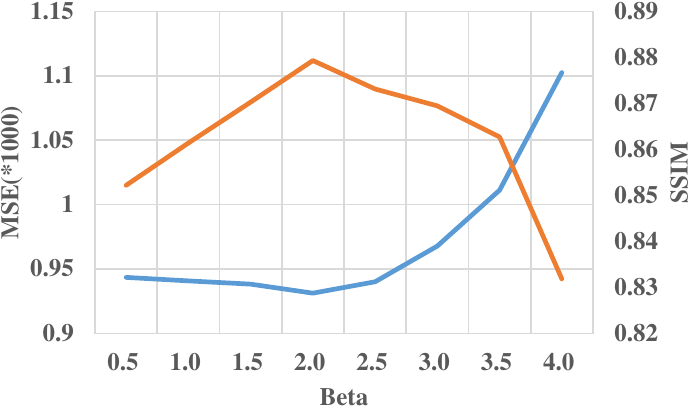}}
\subfigure[]{
\label{Fig10.sub.4}
\includegraphics[width=4.2cm]{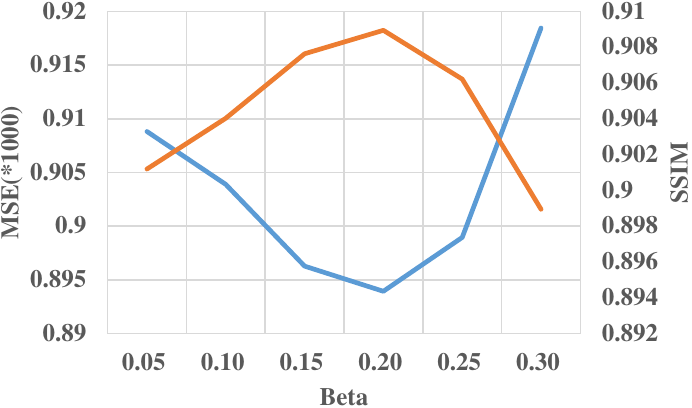}}
\caption{Hyper-parameters selection in comparison methods. (a) Number of iterations for LM-OSEM. (b) Number of iterations for LM-SPDHG. (c) Beta for LM-OSEM-TV. (d) Beta for SPDHG-TV}
\label{hyper}
\end{figure}

Fig. \ref{hyper} shows the selection of hyperparameters for traditional methods. We calculated the MSE and SSIM between images reconstructed with different hyperparameters and the label images to choose the best hyperparameters.

Despite the dynamic access method used in this study enabling on-the-fly calculation of the system matrix for list-mode data and making the deep unrolled method possible on TOF PET list-mode data, the computation of the system matrix for list-mode data still requires high storage space because of its close relationship with the count. This limits the batch size to 1 for experiments with counts of 1e6, resulting in long training times for the network. For instance, training a model with thousands of data points on an NVIDIA TITAN X requires four to five days, which is a current limitation of this method. However, since the trained network can be used directly during testing, the training time is not a major concern. Tab. \ref{time}  shows the required computational time during testing for a single image. All methods were inferred on a GPU, with other traditional methods also benefiting from GPU-accelerated reconstruction. It can be observed that the proposed method demonstrates a competitive reconstruction speed with LM-OSEM, which is practically used in clinical trials.

\section{Conclusion}
In this work, we propose a deep unrolled method for TOF-PET list-mode reconstruction based on learned primal dual. The system matrix for TOF list-mode data is computed on-the-fly using parallel computing acceleration. Both simulation and clinical data show that the proposed method achieves better performance in low-count situations compared to LM-OSEM, LM-EMTV, LM-SPDHG, LM-SPDHG-TV, and FastPET-2D, indicating the promising application of the proposed method for TOF-PET scanners and clinical trials.

\appendices

\section*{Acknowledgment}
We would like to thank Dr. Georg Schramm for sharing the code of PARALLELPROJ in the Github.

\bibliographystyle{IEEEtran}
\bibliography{ref.bib} 

\end{document}